\documentclass[conference]{IEEEtran}
\IEEEoverridecommandlockouts

\usepackage{amsmath,amssymb,amsfonts}
\usepackage{algorithmic}
\usepackage{graphicx}
\usepackage{xcolor}
\usepackage[hidelinks]{hyperref}

\usepackage{dblfloatfix}
\begin{document}

\title{NUMA balancing hampering performance of spiking network simulations}

\author{
\IEEEauthorblockN{Melissa Lober}
\IEEEauthorblockA{\textit{Institute for Advanced Simulation}\\\textit{(IAS-6)} \\
\textit{Jülich Research Centre} \\
\textit{RWTH Aachen University} \\
Jülich/Aachen, Germany \\
m.lober@fz-juelich.de}
\and
\IEEEauthorblockN{Alp Inangu}
\IEEEauthorblockA{\textit{Institute for Advanced Simulation}\\\textit{(IAS-6)} \\
\textit{Jülich Research Centre} \\
\textit{RWTH Aachen University} \\
Jülich/Aachen, Germany}
\and
\IEEEauthorblockN{Gorka Peraza Coppola}
\IEEEauthorblockA{\textit{Institute for Advanced Simulation}\\\textit{(IAS-6)} \\
\textit{Jülich Research Centre} \\
\textit{RWTH Aachen University} \\
Jülich/Aachen, Germany}
\and[\hfill\\\hfill]
\IEEEauthorblockN{Dennis Terhorst}
\IEEEauthorblockA{\textit{Institute for Advanced Simulation}\\\textit{(IAS-6)} \\
\textit{Jülich Research Centre} \\
Jülich, Germany}
\and
\IEEEauthorblockN{Sebastian Gillessen}
\IEEEauthorblockA{\textit{Institute for Advanced Simulation}\\\textit{(IAS-6)} \\
\textit{Jülich Research Centre} \\
Jülich, Germany}
\and
\IEEEauthorblockN{Jan Vogelsang}
\IEEEauthorblockA{\textit{Neuromorphic Software Ecosystems}\\\textit{(PGI-15)}\\
\textit{Jülich Research Centre} \\
Jülich, Germany}
\and[\hfill\\\hfill]
\IEEEauthorblockN{Hans Ekkehard Plesser}
\IEEEauthorblockA{\textit{Department of Data Science,}\\\textit{Faculty of Science and Technology} \\
\textit{Norwegian University of Life Sciences} \\
\textit{Institute for Advanced Simulation}\\\textit{(IAS-6)}\\\textit{Jülich Research Centre} \\
Ås, Norway / Jülich, Germany}
\and
\IEEEauthorblockN{Brian Wylie}
\IEEEauthorblockA{\textit{Jülich Supercomputing Centre} \\
\textit{Jülich Research Centre} \\
Jülich, Germany}
\and
\IEEEauthorblockN{Benedikt Steinbusch}
\IEEEauthorblockA{\textit{Jülich Supercomputing Centre} \\
\textit{Jülich Research Centre} \\
Jülich, Germany}
\and[\hfill\\\hfill]
\IEEEauthorblockN{Guido Trensch}
\IEEEauthorblockA{\textit{Simulation and Data Laboratory Neuroscience} \\
\textit{Jülich Supercomputing Centre}\\\textit{Jülich Research Centre} \\
Jülich, Germany}
\and
\IEEEauthorblockN{Susanne Kunkel}
\IEEEauthorblockA{\textit{Neuromorphic Software Ecosystems}\\\textit{(PGI-15)} \\
\textit{Jülich Research Centre} \\
Jülich, Germany}
\and
\IEEEauthorblockN{Markus Diesmann}
\IEEEauthorblockA{\textit{Institute for Advanced Simulation}\\\textit{(IAS-6)}\\\textit{Jülich Research Centre} \\
\textit{Department of Physics, Faculty 1}\\
\textit{Department of Psychiatry,Psychotherapy}\\\textit{and Psychosomatics}\\\textit{School of Medicine}\\\textit{RWTH Aachen University} \\
Jülich/Aachen, Germany}
}

\maketitle

\begin{abstract} 
Computing centers today mostly operate conventional CPU- and GPU-based systems, where the direct way of decreasing energy consumption is a reduction in the applications' runtime.
Neuromorphic computing promises an alternative architecture with improved energy efficiency for artificial intelligence. In this endeavor, code for the simulation of large-scale spiking networks on conventional supercomputers is the reference.
We show that turning off automatic NUMA balancing may reduce energy consumption by $\mathbf{30\,}$\textbf{\%}. This dwarfs other attempts of increasing the energy efficiency of a computing center with respect to cost effectiveness.
The memory access pattern of spiking network simulation code dynamically interacts with automatic NUMA balancing. This does not affect the correctness of simulation results and thus goes unnoticed in day-to-day neuroscience research. In performance analysis, however, time measurements fluctuate obstructing attempts to optimize simulation technology.
A new time- and compute-node resolved performance display exposes the fine-grained temporal variability in the course of distributed spiking network simulations. The analysis uncovers that automatic NUMA balancing is of disadvantage and, in particular, affects the jemalloc library for thread-aware memory allocation in a transient manner. The method also allows developers to detect perturbations of the HPC system and target specific improvements to simulation technology.
As a consequence of these findings, we have equipped our supercomputers with an option to turn on or off automatic NUMA balancing on a per-job basis on the user level. This gives researchers the opportunity to find the best setting for the application at hand. There are indications in the literature that the effect has been observed before, yet it does not seem common knowledge in scientific computing. It remains to be investigated how widespread the phenomenon is among scientific codes.
\end{abstract}

\section{Introduction}
\label{sec:Introduction}
There is a discrepancy of orders of magnitude between the energy consumption of current supercomputers and the human brain. While the hardware substrate is different, the two systems are also optimized for different activation patterns: repetitive dense matrix multiplications on the one side, and sparse spatio-temporal activity on the other. Therefore, the field of neuromorphic computing strives to harness the architectural differences for a reduction of the energy consumption of artificial intelligence.
Any relevant neuromorphic design needs to outperform conventional computing systems in either speed, energy consumption, or both. This requires the availability of portable benchmarks and highly optimized simulation code exposing the real hardware limits of conventional systems. A model of the cortical microcircuit \cite{Potjans14_785} became a de facto standard benchmark, and a range of neuromorphic systems pass the benchmark today faster than real time \cite{Senk26}. This benchmark constitutes a relevant milestone for the field as any larger model of the mammalian cortex necessarily has a lower average connection probability and should therefore be easier to simulate. Research is now ongoing towards next generation benchmarks. A promising target is a multi-area model (MAM, \cite{Schmidt18_e1006359,Schmidt18_1409}), demonstrating the interaction of local and global connectivity. The MAM challenges conventional supercomputers because of its communication demands and neuromorphic systems due to its sheer size. Nevertheless, first results based on GPUs as a bridge technology are coming up \cite{Tiddia22_883333,Golosio26_024012,Knight2021_136,Knight26}, and new large-scale neuromorphic systems are under construction \cite{Kudithipudi25}.

\begin{figure}
    \centering
    \includegraphics[width=0.5\linewidth]{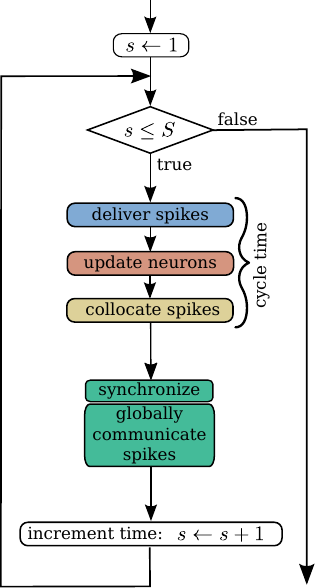}
    \caption{\textbf{Simulation cycle of distributed spiking network code}. Colors indicate simulation phases; brace defines the phases contributing to the simulation-cycle time.}
    \label{fig:propagation_cycle}
\end{figure}

The distributed simulation code NEST \cite{Gewaltig_07_11204} for spiking neuronal networks serves as a conventional CPU-based reference for new systems. In every simulation cycle, each MPI process independently performs spike delivery to local target neurons, update of all local neurons, and collocation of locally generated spikes; each cycle terminates with global synchronization and spike communication between MPI processes (Fig.~\ref{fig:propagation_cycle}). Spike delivery and neuron update are fully thread-parallel, while collocation and communication are executed by the master thread alone. The global synchronization requires all MPI processes to wait for the slowest participating process. The real-time factor (RTF), defined as the ratio of wall-clock time to simulated biological time, therefore accumulates the slowest simulation-cycle times across all processes and all simulation cycles \cite{Lober26}.

On conventional supercomputers, neither bandwidth nor latency limit simulations of the MAM but the variability in simulation-cycle times, which translates into long synchronization times \cite{Lober26}. However, the cycle time only weakly depends on computational load such that an imbalance of load does not explain it. This situation motivates us to investigate how much of the variability comes from outside the simulation code and is due to the operating system or the hardware.
One such operating system feature is automatic NUMA balancing, which was introduced with full scheduler support in Linux 3.13 \cite{Corbet13,Gorman13}. On multi-socket servers, memory attached to a remote processor socket incurs higher access latency than local memory. To reduce remote memory accesses, the kernel periodically unmaps memory pages and records which NUMA domain faults on each page, then migrates pages toward the domain that accesses them most often. The kernel documentation notes that the overhead of unmapping and fault handling does not always yield a net performance improvement \cite{LinuxKernelDoc}.

In the following we first show that NUMA balancing interferes with the distributed and parallel operation of the NEST simulation code in the presence of thread aware memory allocation. We then confirm that the phenomenon also occurs for the system allocator, and finally that it is not particular to a specific number of compute nodes.

The presented conceptual and algorithmic work is part of our long-term collaborative project to provide the technology for neural systems simulations \cite{Gewaltig_07_11204}.
Preliminary results have been presented in abstract form \cite{Lober26-NESTConf,Lober26-ICNCE}.

\section{Results}
\label{sec:Results}

\begin{figure*} 
    \centering
    \includegraphics[width=1\textwidth]{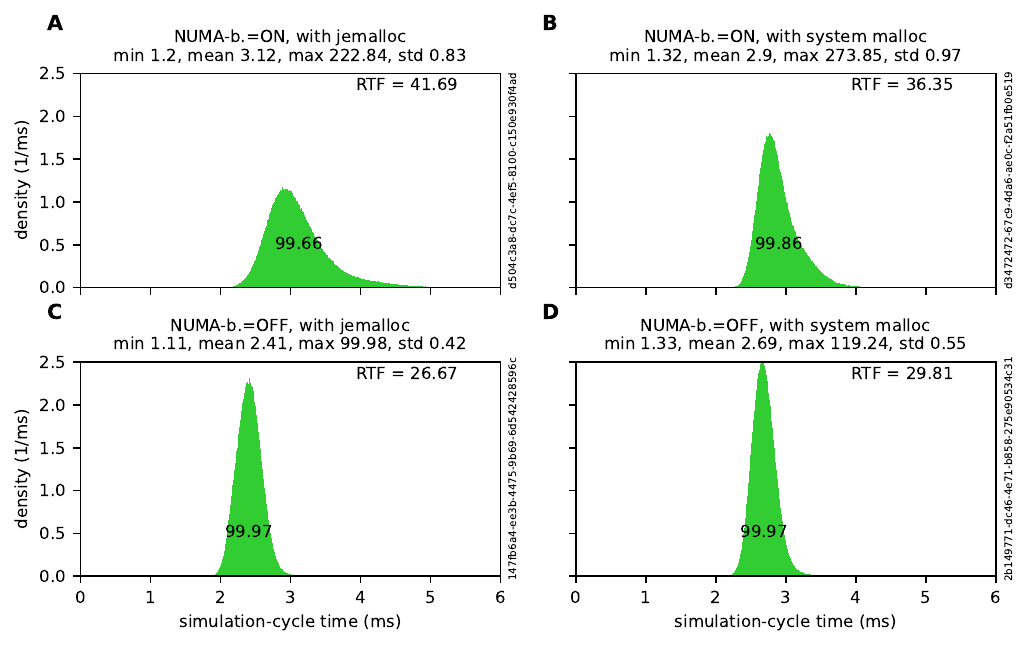}
    \caption{\textbf{Distribution of cycle times pooled across all MPI processes and simulation cycles}. Cycle times in the presence of automatic NUMA balancing (panels \textbf{A, B}) and without (\textbf{C, D}) using jemalloc (\textbf{A, C}) or the system allocator (\textbf{B, D}). Data shows results of simulations of a multi-area cortical network model (MAM, \protect\cite{Schmidt18_e1006359,Schmidt18_1409}) for $10$ seconds of biological time using $16$ MPI processes with $64$ threads each on $8$ compute nodes of JURECA-DC. Panels display density only for cycle times up to $6$ milliseconds (value in distribution shows \% covered), and titles show mean and standard deviation of the distributions, and the maximum cycle time observed in milliseconds. Superimposed values in the top right corners show ratio of wall-clock time and full stretch of simulated biological time (realtime factor, RTF). Vertical text in right margin of plots reading bottom to top uniquely identifies the data set (UUID).}
    \label{fig:histograms_jemalloc_on}
\end{figure*}
Fig.~\ref{fig:histograms_jemalloc_on} shows the distribution of simulation-cycle times pooled across all MPI processes and cycles of a MAM simulation with NEST for four combinations of NUMA-balancing state and memory allocator. With automatic NUMA balancing enabled and jemalloc \cite{Evans06} as the memory allocator (Fig.~\ref{fig:histograms_jemalloc_on}A), the distribution exhibits a pronounced long tail extending to cycle times well above the mean, with a maximum observed cycle time of $222.84\,\text{ms}$ and a standard deviation of $0.83\,\text{ms}$. Because the RTF is set by the maximum cycle time encountered across all processes at each cycle, this long tail disproportionately inflates the total wall-clock time. 
With NUMA balancing disabled and jemalloc active (panel C), the long tail in the cycle time distribution is eliminated: the maximum observed cycle time drops to $99.98\,\text{ms}$ and the distribution becomes concentrated around the mean with the standard deviation narrowing to $0.42\,\text{ms}$.

\begin{figure*} 
    \centering
    \includegraphics[width=1\textwidth]{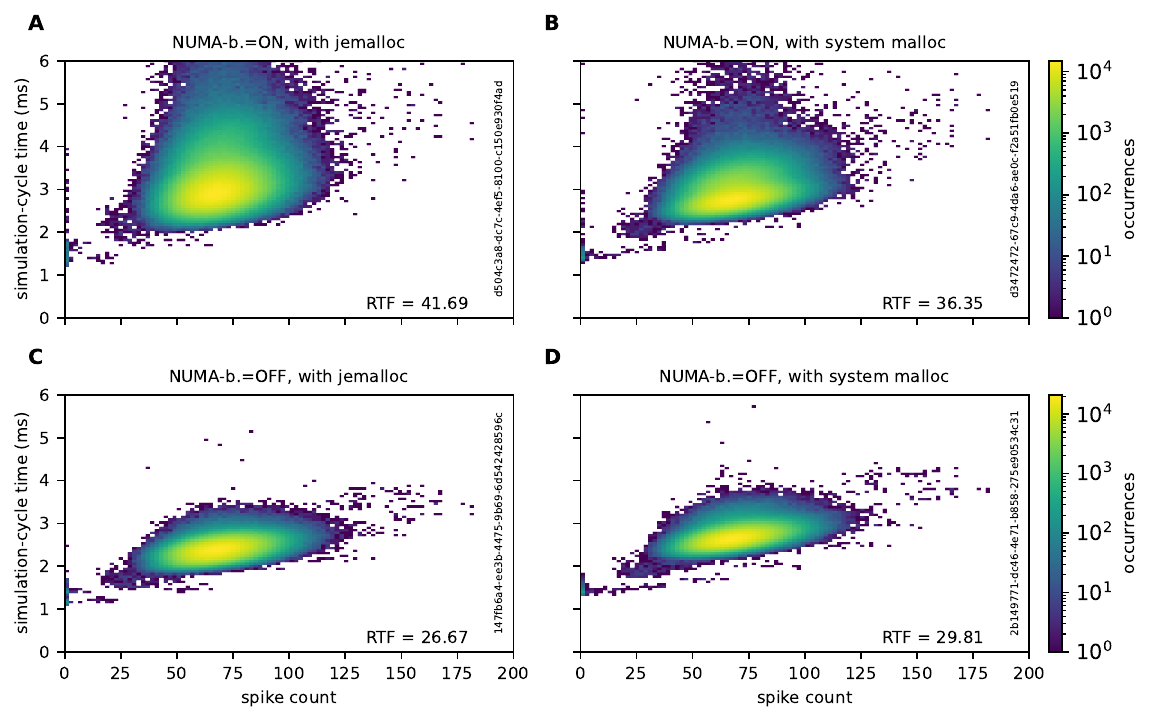}
    \caption{ \textbf{Correlation of cycle time to spike count.} Same arrangement of panels and annotation as in Fig.~\ref{fig:histograms_jemalloc_on}. Data pooled over all MPI processes in histogram with 100 bins along each dimension (no smoothing). Counts normalized to the color range of the perceptually uniform sequential map 'viridis' using the logarithmic transform LogNorm of Matplotlib \cite{Hunter07_visuzalization}. White means no counts in observation interval.}
    \label{fig:correlation_spike_count}
\end{figure*}
A natural hypothesis is that the long-tail cycle times in Fig.~\ref{fig:histograms_jemalloc_on}A reflect periods of elevated network activity, as higher spike counts increase the computational load of spike delivery and collocation. Fig.~\ref{fig:correlation_spike_count} tests this hypothesis by plotting cycle time against the spike count of the previous cycle for each MPI process. With NUMA balancing enabled (panel A), the longest cycle times show no correlation with spike count. This rules out network dynamics as the cause of the long tail. The effect therefore originates at the machine level, outside the simulation code itself. 
In the absence of NUMA balancing, a weak dependence of cycle time on spike count emerges (panel C). While spike count varies by a factor of five, cycle time increases by only $50\%$ over this interval. Assuming a linear relationship between cycle time and spike count, the squared Pearson correlation coefficient indicates that spike count accounts for only $17\,\%$ of the total variability in cycle time. The residual variability of cycle time, however, is independent of spike count and of the same order as the mean such that a broad slightly tilted ellipse dominates the graphs. We recognize this structure also in the presence of NUMA balancing (panel A) for the low cycle times. Nevertheless, a complex additional cloud emerges reaching out to cycle times of three times the mean with decreasing probability. The cloud appears as composed of further ellipses like the main one centered at progressively larger means. The stack of ellipses fuses into a vertical band destroying the correlation. 

\begin{figure*} 
    \centering
    \includegraphics[width=1\textwidth]{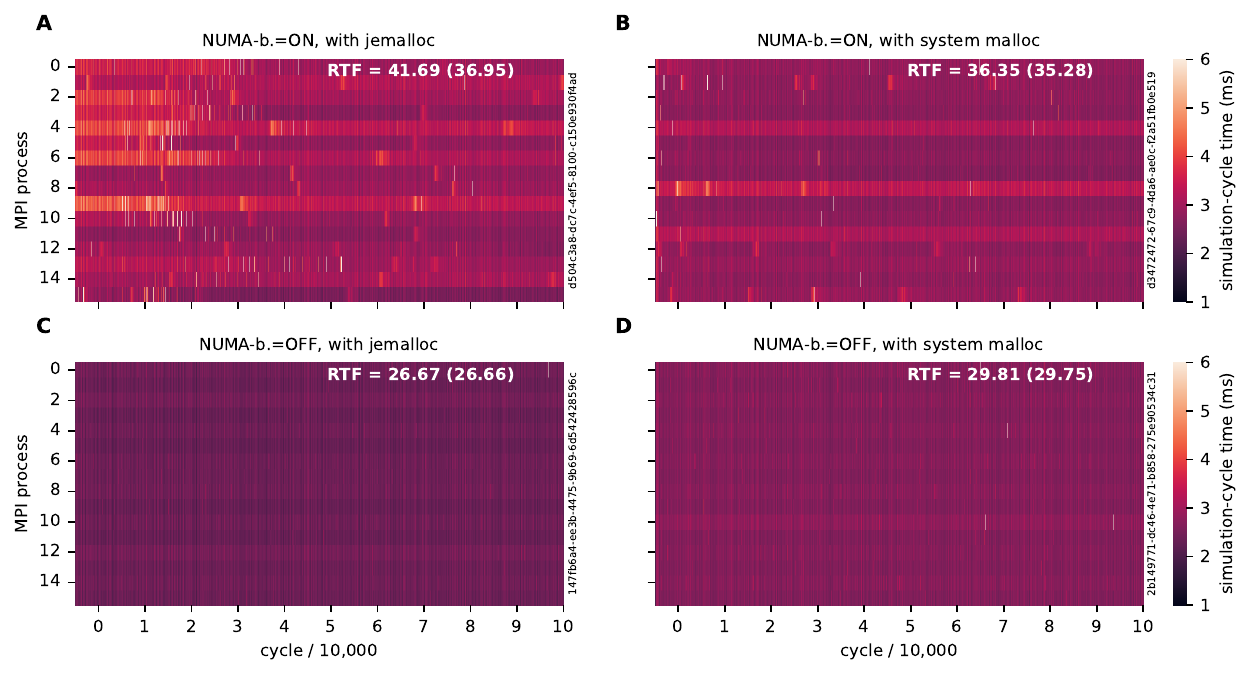}
    \caption{\textbf{Simulation-cycle time resolved by MPI process and simulation cycle}. Cycle times are color coded. Same arrangement of panels and annotation as in Fig.~\ref{fig:histograms_jemalloc_on}. Resolution of horizontal axis in simulation cycles ($0.1\,\text{ms}$ model time). Pairs of consecutive MPI ranks starting with $0$ (vertical) are on same node. Cutoff of cycle time is at $6\,\text{ms}$, all larger values indicated by white. RTF in parentheses refers to second half of simulation.}
    \label{fig:heatmap_jemalloc_on}
\end{figure*}
Fig.~\ref{fig:heatmap_jemalloc_on} shows cycle times as a function of both simulation cycles and MPI process, providing time- and process-resolved insight into the origin of the long tail. With NUMA balancing enabled and jemalloc active (panel A), four characteristic patterns are visible. First, cycle times are elevated across nearly all processes during the initial thousands of simulation cycles before gradually subsiding. This phenomenon originates from the Linux kernel's adaptive NUMA balancing heuristic, as sketched below.
Second, even after the transient subsides, a subset of processes exhibits persistently elevated cycle times relative to their peers throughout the remainder of the simulation, producing a two-tone horizontally striped pattern in the heatmap.
A third pattern is vertical stripes spanning all processes. These stripes reflect the fluctuations in neuronal activity, where elevated activity causes longer cycle times. The stripes extend across all processes because the round-robin distribution of neurons over processes homogenizes computational load. Hence, even a fluctuation confined to a single brain area affects all processes equally. Finally, there are occasional cycles in individual processes exhibiting extremely long cycle times. These blips carry a vanishing probability mass, and therefore their contribution to total wall-clock time of the simulation is insignificant. 
With NUMA balancing disabled, the heatmaps show that both the initial transient and the striped process asymmetry are largely absent and the blips are even more rare (panel C). The resulting reduction in RTF amounts to approximately $30\,\%$. There is a faint horizontal two-tone pattern of stripes in the background. This indicates that the two MPI processes running on a node do not find exactly identical conditions. As the MPI processes use all cores but share the operating system, one of them may be affected by system tasks. Even for the second half of the simulation, disabling NUMA balancing reduces the RTF by $28\,\%$. 

\begin{figure}[h]
    \centering
    \includegraphics[width=1\linewidth]{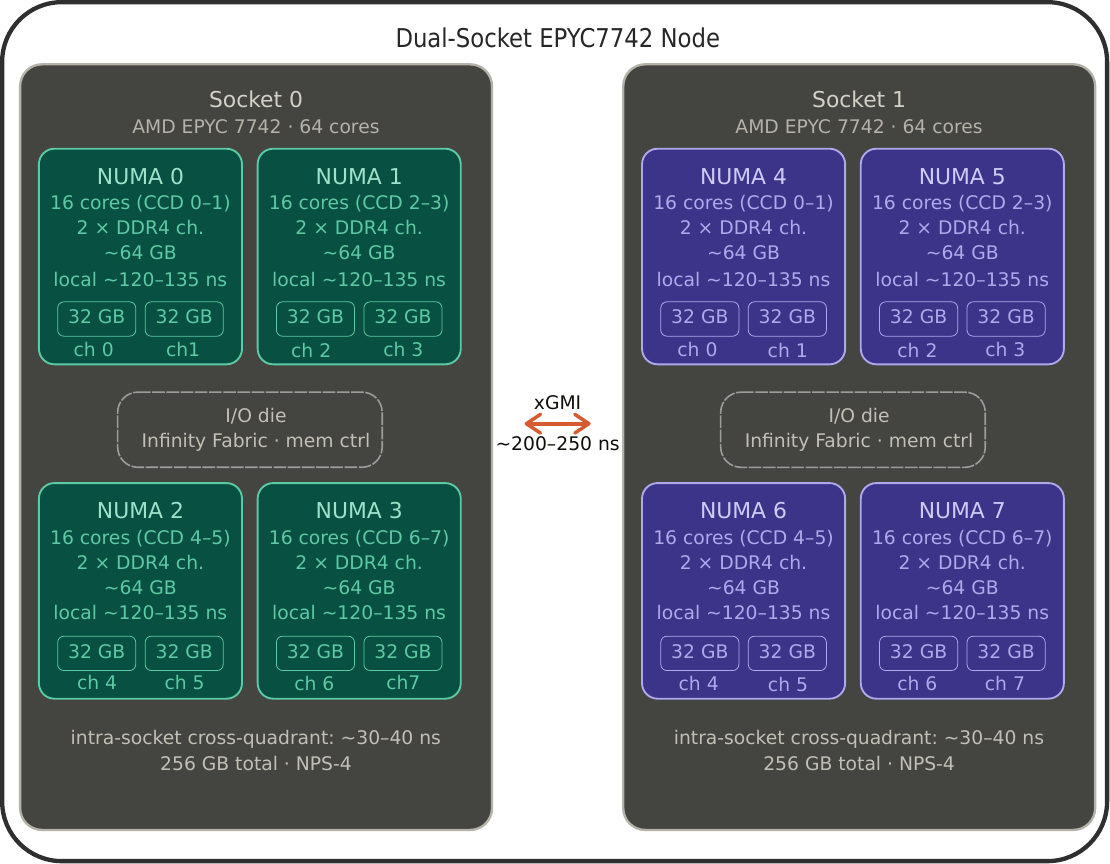}
    \caption{\textbf{Schematic of a JURECA-DC compute node} (Atos BullSequana X$2410$) comprising two AMD EPYC $7742$ processors. Each socket contains $64$ cores organised into eight Core Complex Dies (CCDs), grouped into four I/O-die quadrants. The BIOS is configured at NPS-4 (Non-Uniform Memory Access, $4$ domains per socket), exposing eight NUMA domains per node. Each domain spans two CCDs ($16$ cores) and two DDR4-3200 memory channels, each populated with one $32$ GB DIMM, giving $64$ GB of local memory per domain and $512\,$GB per node in total. The two sockets are connected via AMD's inter-socket xGMI link. Memory access latency increases with topological distance: from approximately $120–135\,$ns within a domain, to approximately $30–40\,$ns across quadrants within the same socket, to approximately $200–250\,$ns across the xGMI link.}
    \label{fig:jureca_node_architecture}
\end{figure}

JURECA-DC compute nodes are configured at NPS-4 (Fig.~\ref{fig:jureca_node_architecture}), exposing eight NUMA domains per node in total and four per socket. In our simulations, one MPI process is placed on each socket, with $64$ threads explicitly pinned one-to-one to the $64$ physical cores of that socket, fully occupying all four NUMA domains. The Linux kernel's automatic NUMA balancing mechanism continuously monitors memory access patterns and attempts to migrate memory pages toward the NUMA domain of the core most frequently accessing them, with the goal of reducing remote memory access latency. With four NUMA domains per socket, each hosting $16$ of the $64$ pinned threads, the balancer has three potential migration targets within the socket alone for any given page.

\begin{figure}[h]
    \centering
    \includegraphics[width=1\linewidth]{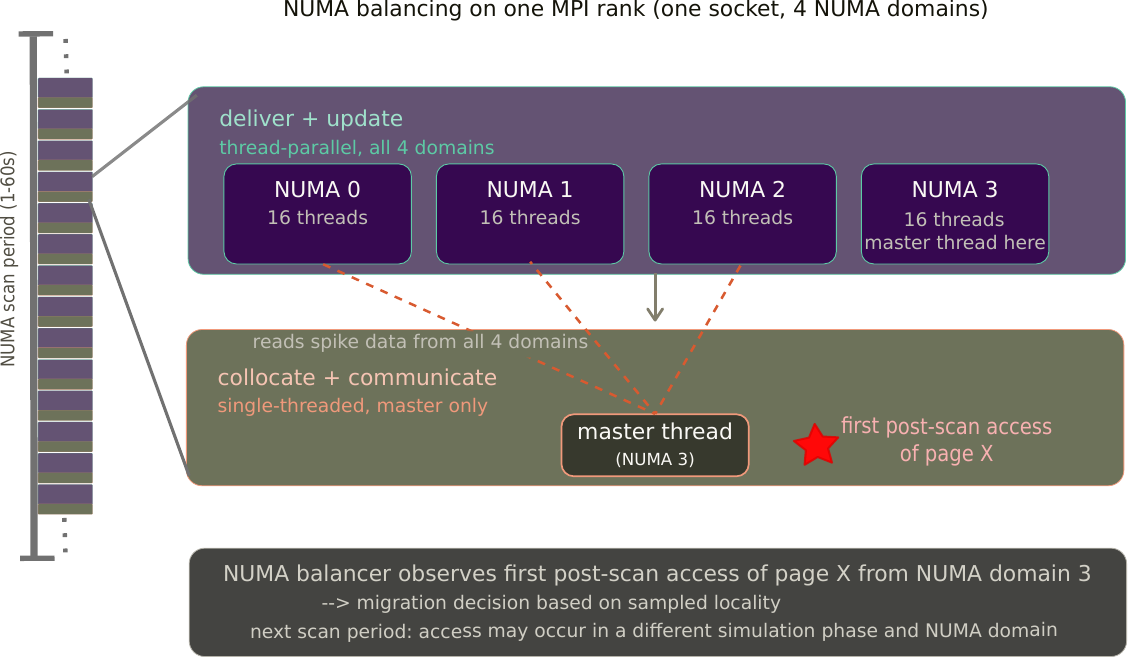}
    \caption{\textbf{Schematic of the interaction between the simulation cycles and automatic NUMA balancing}, as executed by a single MPI process occupying one full socket (four NUMA domains, $64$ cores). The deliver and update phase (top aubergine block) is fully thread-parallel, with $16$ threads executing in each of the four NUMA domains. The subsequent collocation and communication phase (center spinach block) is single-threaded, executed by the master thread residing in one domain (here NUMA 3), which reads spike data written by threads across all four domains. The scan period of one to sixty seconds (indicated by gray interval on left) spans thousands of simulation cycles each lasting a few milliseconds of wall-clock time (zoom-out from blocks), during which at most one fault per page is recorded. The balancer attributes the fault to whichever thread happens to access the page first after the unmap event and issues a migration decision accordingly. In this example, the master thread residing on NUMA domain 3 happens to access memory page X first post scan (red star).
}
    \label{fig:numa_during_cycle}
\end{figure}

Fig.~\ref{fig:numa_during_cycle} illustrates how the execution structure of NEST interacts with the Linux kernel's automatic NUMA balancing on JURECA-DC. In NEST's simulation cycle, spike delivery and neuron update are fully thread-parallel, with threads distributed across all four NUMA domains of the socket. The subsequent collocation and global communication of spikes, however, are single-threaded and executed exclusively by the master thread, which resides in one specific NUMA domain determined by its core assignment. During collocation and communication, the master thread --- the single thread executing these serial phases while all other threads are idle --- reads spike-buffer data that was written in parallel by all threads across the four NUMA domains of the socket. 

Automatic NUMA balancing periodically replaces page mappings with NUMA hinting entries. The first subsequent access to such a page generates a NUMA hinting fault, allowing the kernel to record the accessing task and NUMA node before restoring the original mapping. These sparse observations contribute to the locality statistics used to guide page migration decisions. The scan period ranges from one to sixty seconds, while one simulation cycle takes only a few milliseconds of wall-clock time. Therefore, thousands of simulation cycles elapse before the kernel decides whether pages should be migrated.

Since the simulation repeatedly alternates between a fully parallel phase (delivery and update) and a single-threaded phase (collocation and communication), and both phases access a substantial fraction of the memory, the simulation phase in which the first post-scan access to a page occurs becomes largely coincidental. Consequently, the observation represents only a single access, while many subsequent accesses to the same page remain unobserved until the page is selected for hinting again. A possible scenario is the following: The locality statistics used by the NUMA balancer reflects the simulation phase in which pages happened to be sampled rather than the phase that dominates execution time. In this situation the kernel performs page migrations that provide little lasting benefit while still incurring the runtime overhead of automatic NUMA balancing. The transient slowness observed in Fig.~\ref{fig:heatmap_jemalloc_on}A is consistent with the adaptive scan period of NUMA balancing, which initially scans memory more frequently before progressively reducing scan frequency. 
However, the persisting cycle-time asymmetry across processes in the heatmaps (Fig.~\ref{fig:heatmap_jemalloc_on}A,B) remains to be fully understood. 

As a consequence of the analysis above, we introduce a slurm \cite{Jette03} option allowing users to enable or disable automatic NUMA balancing on a per-node basis at the job level, without requiring system administrator intervention (see \ref{sec:Methods} for details). 

We next investigate whether the phenomenon is particular to the use of the custom memory allocator jemalloc instead of the system allocator. A thread aware allocator is relevant for the simulation code because network construction (not discussed here) needs to be a parallel activity \cite{Ippen2017_30}. Using the system allocator instead of jemalloc in the presence of NUMA balancing still leads to an asymmetric cycle-time distribution but its tail is less pronounced and the mean is lower (Fig.~\ref{fig:histograms_jemalloc_on}B). The respective heatmap Fig.~\ref{fig:heatmap_jemalloc_on}B uncovers that the initial slowness disappears, reducing the overall RTF, but the horizontal stripes persist. They appear slightly fainter than in the presence of jemalloc, an impression confirmed by a small reduction in second-half RTF. The structure of the correlation graph remains essentially unaltered (Fig.~\ref{fig:correlation_spike_count}B).

As previously observed for jemalloc, turning off NUMA balancing also removes the tail of the cycle-time distribution when using the system allocator (Fig.~\ref{fig:histograms_jemalloc_on}D). The correlation graphs (Figs.~\ref{fig:correlation_spike_count}C and D) as well as the heatmaps (Figs.~\ref{fig:heatmap_jemalloc_on}C and D) show similar trends for jemalloc and the system allocator. However, while disabling NUMA balancing substantially reduces overall and second-half RTF for both allocators, jemalloc consistently achieves the lower cycle times. Thus, while  NUMA balancing perturbs the memory placement achieved by jemalloc more severely than that achieved by the system allocator, the best configuration uses jemalloc and abstains from NUMA balancing.

\begin{figure}
    \centering
    \includegraphics[width=0.45\textwidth]{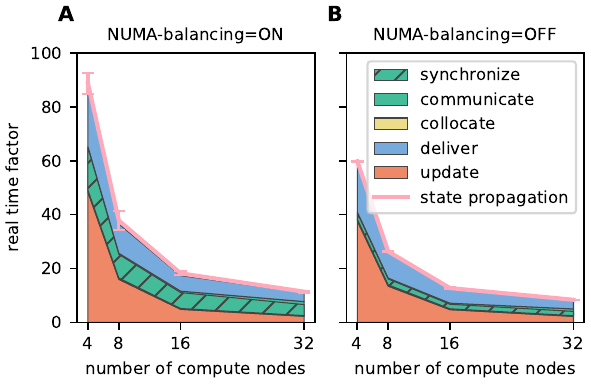}
    \caption{\textbf{Strong-scaling benchmark of the MAM.} \textbf{A}: Automatic NUMA balancing ON; \textbf{B}: automatic NUMA balancing OFF, in presence of jemalloc for ground state of network dynamics. Threads and two MPI processes per node as in Fig.~\ref{fig:histograms_jemalloc_on}. Color code of stages of the update cycle (legend) corresponds to Fig.~\ref{fig:propagation_cycle}. State propagation (pink) indicates contributions to total simulation time not accounted for by sum of stages. Error bars indicate standard deviation over three repetitions (collapsing to horizontal bar in \textbf{B}). 
    \label{fig:MAM_strong-scaling}}
\end{figure}

To assess whether the phenomenon is specific to a particular problem size or a single random instantiation of the network, Fig.~\ref{fig:MAM_strong-scaling} shows strong-scaling results for the MAM with NUMA balancing enabled and disabled across four node counts and three independent random seeds. With NUMA balancing enabled, the RTF is consistently elevated relative to the NUMA-balancing-off condition across all node counts and seeds tested, and the variability between seeds is substantially larger. The effect is therefore a systematic property of the interaction between the code's execution pattern and the NUMA balancing heuristic.

\section{Discussion}
\label{sec:Discussion}

The research described here was carried out over a period of more than a year, from the formalization of the phenomenon into a ticket (J\"ulich Supercomputing Center JSC ticket \#10100311) to the solution alone nine month were required (April to December 2025). The starting point were unexplainable fluctuations in simulation times in a project on the optimization of the mapping of a brain-scale neuronal network onto a supercomputer \cite{Lober26}. Initially, we gathered confidence that the effect originates from outside our code by control simulations on similar hardware with an independent configuration and software stack. 

The five main findings of our study are:
\begin{itemize}
    \item automatic NUMA balancing may interact badly with a memory access pattern like the one of the NEST code
    \item implementation of timers in the simulation code NEST (from release 3.10) reporting the duration of individual cycles per node     
    \item a new diagram (heatmap) collocating the cycle times of all nodes as a function of consecutive simulation cycles 
    \item a new diagram showing the variability of cycle times as a function of the workload
    \item a software switch for the queueing system slurm exposing automatic NUMA balancing to the user-level
\end{itemize}

There are clear statements in the literature that automatic NUMA balancing can degrade the performance of HPC applications \cite{Gaud15,AMD23,GormanJambor24}. However, the NUMA nodes per socket (NPS) setting and automatic NUMA balancing are largely treated independently. Research is going on to further improve the algorithms for automatic balancing, see e.g.~\cite{Gaud23,LiuYu24}. At least in the field of computational neuroscience, there is little awareness of the potential for large gains in energy efficiency by optimizing such settings and, vice versa, the potential for catastrophic performance if they are overlooked.

The neuronal network model we use in the present study is well established and has produced neuroscientific insight \cite{Schmidt18_e1006359}. At the same time it is still at the edge of what neuromorphic computing systems can cope with today in terms of sheer memory requirements but also communication load~\cite{Tiddia22_883333,Golosio26_024012,Knight2021_136,Knight26}. In some of the brain areas of the network model, the synchronization of the neuronal activity is higher than in nature. This stresses any computing system more than necessary. Improved brain-scale models may ease this stress on communication.

Looking back, we have carried out thousands of neuroscientific production runs since the inception of the JURECA-DC supercomputer in 2020. As the simulation results were always correct and arrived comfortably fast, the performance issue went unnoticed for a long time. Only in a technical project aiming at the improvement of the communication architecture of NEST for brain-scale models we became suspicious about the fluctuations in run time.

It remains an open question how wide spread the phenomenon is among scientific codes. The communication pattern of NEST characterized by frequent collective communication between nodes with symmetrical parallel action of all threads in between is certainly extreme. Nevertheless, the opportunity to reduce energy costs of calculations by $30\,\%$ just by improving the software stack, in this case by a single switch, seems relevant as other measures, like the improvement of computing hardware or cooling systems, may be more costly.

The combination of a supercomputer constructed for research at the frontier of what is possible today and a research code that is continuously in flux is difficult to debug. The present work is not a tutorial on how to track down a problem in this situation, but describes a particular phenomenon using a single system. Briefly, in the best case the problem is clearly on the side of the computing system or on the side of the research code; but which side is it? In our experience it is helpful to maintain access to two supercomputers with independent software stacks despite the additional administrative efforts. As a further measure we suggest that research groups inspect a new system with time-resolved instrumentation of their real-world application code before production starts. For a generic simulation code like NEST serving a range of different projects the effort is manageable as the validation has to be done only once. This underlines the necessity to consider relevant research software as scientific infrastructure \cite{Hocquet24_1_tmp}.

Over a period of several months the consortium has systematically investigated the effects of settings on all levels from hardware configuration to application code. This required a robust collaborative workflow and the built-up of a data base to structure the communication between researchers in different units and of different level of experience. A recently developed benchmarking framework \cite{Vogelsang26_arxiv} supported the investigation and the experience gathered flows back into the refinement of such concepts.

Even in the absence of automatic NUMA balancing, the distribution of simulation-cycle times remains on the order of the mean cycle time and only weekly correlates with workload. As the system needs to wait for the slowest node in every communication step, reducing the variability is an effective way of reducing simulation time. Due to this maximum operation even algorithms that slightly increase mean cycle time while reducing variability may be successful.
Further work now needs to consider that on modern compute nodes the simulation code is hybrid: message-passing between nodes but massively multi-threaded within nodes. 
The nodes require a similar $\text{max}$-operation between threads as between nodes. Hence, strategies like sparing a core for the operating system may slow down a compute node in the completion of a cycle but reduce variability.
Understanding whether the variability originates from multi-threading and how it can be reduced demands more fine grained tools and modeling as we employ in the present study.

From the computational neuroscience perspective, our study shows that the performance of our simulation technology is still far away from the limits imposed by hardware. Thus we expect that there is room for further improvements. From the angle of science management the study shows how a team of domain experts, HPC experts, and supercomputer operators can solve a problem with likely impact beyond the domain. Decisive ingredients for the success was a collaborative benchmarking workflow but also mutual trust and flexibility on all sides.

\section{Methods}
\label{sec:Methods}

\subsection*{Simulation engine}
All benchmarks are performed with NEST~3.10 \cite{Nest310}, an open-source simulation code for large-scale spiking neuronal networks. NEST uses hybrid parallelization based on MPI for distributed-memory execution and OpenMP for shared-memory parallelism. During each simulation cycle, spikes are delivered, neuronal states are updated, and newly generated spikes are collected before a global communication step exchanges spikes between MPI processes to maintain causality (Fig.~\ref{fig:propagation_cycle}).
To evaluate performance, we use NEST's built-in high-resolution timers to measure the runtime of the individual simulation phases (see next paragraph for details). 
Reported runtimes are averaged across all MPI processes, and performance is expressed as the real-time factor, defined as the wall-clock time normalized by the simulated model time. For the simulation of the MAM, one simulation cycle advances the model by $0.1\,\text{ms}$.

\subsection*{Cycle-resolved timers}
Starting with NEST~3.10, the simulation code can be configured (cmake option \texttt{-Dwith-cycle-timers=ON}) to record per-cycle timing information and spike counts, a profiling capability introduced as part of the present work (see Figs.~\ref{fig:histograms_jemalloc_on}-\ref{fig:heatmap_jemalloc_on}).
Since spike communication cannot begin until all MPI processes have completed their local computations, NEST~3.10 also provides the option to separately quantify synchronization and communication time by inserting an explicit \texttt{MPI\_Barrier} immediately before the collective \texttt{MPI\_Alltoall()} operation. This feature is enabled when compiling the code with the cmake option \texttt{-Dwith-mpi-sync-timer=ON}.
More information on the built-in timers of NEST can be found in the user-level documentation\footnote{\url{https://nest-simulator.readthedocs.io/en/stable/index.html}}.

\subsection*{Network model}
All simulations throughout this study use the multi-area model of the macaque visual cortex (MAM\footnote{https://github.com/INM-6/multi-area-model}; \cite{Schmidt18_e1006359,Schmidt18_1409}), a large-scale spiking network model comprising one square millimeter of each of the $32$ visual cortical areas. Each area is represented by a layered microcircuit of excitatory and inhibitory leaky integrate-and-fire neurons \cite{Potjans14_785}. Experimental anatomical data determine area-specific neuron numbers and connectivity, resulting in heterogeneous network structure and activity across cortical areas. Approximately one third of all synapses are long-range projections between areas, with transmission delays sampled from Gaussian distributions. All benchmarks use the model's asynchronous ground-state regime, characterized by stable low-rate activity with an average firing rate of approximately $2.5\,\text{spikes/s}$. The shortest synaptic delay in the model is $0.1\,\text{ms}$.

\subsection*{Benchmarking system}
We perform benchmarks on the CPU partition (JURECA-DC Phase 2) of Jülich Research Centre's JURECA system \cite{Thoernig21_182}. The node architecture is shown in Fig.~\ref{fig:jureca_node_architecture}. Each standard compute node is equipped with two 64-core AMD EPYC $7742$ processors ($128$ cores total) at ($2.25\,\text{GHz}$) and ($512\,\text{GiB}$) of memory. Nodes are interconnected via Mellanox HDR100 InfiniBand. Applications are compiled with GCC and use OpenMPI for distributed-memory communication, OpenMP for shared-memory parallelism, and jemalloc 5.3.0 as the memory allocator. We employ two MPI processes per node, each running $64$ OpenMP threads. Each MPI process is pinned to the physical cores of a single CPU socket, ensuring that the two processes occupy disjoint sets of cores (\texttt{OMP\_PROC\_BIND=close}, \texttt{OMP\_PLACES=threads}, where the application is scheduled with \texttt{srun --cpubind=mask\_cpu:0xFFFFFFFFFFFFFFFF,\linebreak[0]{}0xFFFFFFFFFFFFFFFF0000000000000000}). Simultaneous multithreading (SMT) is disabled.
Benchmarks are performed using the automated benchmarking pipeline \textit{CI-beNNch} \cite{Vogelsang26_arxiv} that utilizes continuous integration principles to enable fully automated, reproducible, and user-independent benchmarking workflows.

\subsection*{User-level control of automatic NUMA balancing}
As a consequence of the findings of this study, we have equipped our supercomputers, including JURECA-DC, with a NUMA balancing switch accessible at the user level via the slurm job scheduler. Automatic NUMA balancing is enabled by default and can be disabled for a given job by adding \texttt{\#SBATCH --numa-balancing=0} to the job script.

\section*{Acknowledgments}
This project received funding from NeuroSys as part of the initiative “Clusters4Future” funded by the Federal Ministry of Education and Research BMBF (03ZU1106CB,03ZU2106CB); the German Research Foundation (DFG) - 368482240/GRK2416 (MultiSenses-MultiScales) and 545776403/FOR5880 (Mod4Comp); Joint Lab HiRSE, the Helmholtz Platform for Research Software Engineering - an innovation pool project of the Helmholtz Association; the European Union’s Horizon Europe Programme under the Specific Grant Agreement No. 101147319 (EBRAINS 2.0 Project); the Joint Lab ‘Supercomputing and Modeling for the Human Brain’ (SMHB) of the Helmholtz Association; the Volkswagen Foundation - 0071143 (STRUCTICITY); Melissa Lober received funding from the HIDA-NORA Mobility Program for a research stay at NMBU, Norway.

We gratefully acknowledge the Gauss Centre for Supercomputing e.V. (GCS) for funding this project by providing computing time on the supercomputers SuperMUC-NG at Leibniz Supercomputing Centre (LRZ) and HPE Apollo (Hawk) at the High Performance Computing Center Stuttgart (HLRS), and computing time granted by the JARA Vergabegremium and provided on the JARA Partition part of the supercomputer JURECA at Forschungszentrum Jülich (computation grant JINB33).

We thank the members of the NEST Initiative for providing a forum for discussion and continuous support.

\section*{Conflict of Interest Statement} 
The authors declare that the research was conducted in the absence of any commercial or financial relationships that could be construed as a potential conflict of interest.


\section*{Data availability statement}
Source code, simulation and analysis scripts are openly available at Zenodo \cite{Lober26_numbalazenodo}.

\newpage
\bibliographystyle{IEEEtran}
\bibliography{bib/journal_macros_abbreviated,IEEEabrv,bib/brain,bib/references}

\end{document}